\documentclass[letterpaper,11pt]{article}
\usepackage{graphicx}
\usepackage{hyperref}

\title{\large\bf RawArray: A Simple, Fast, and Extensible Archival Format for Numeric Data}
\author{David S. Smith\footnote{\texttt{david.smith@gmail.com}, GitHub: \url{https://github.com/davidssmith}}~\footnote{This work was funded by NCI/NIH K25 CA176219.}}

\date{\it Institute of Imaging Science, Vanderbilt University Medical Center, Nashville, TN, USA}

\begin{document}

\maketitle

\begin{abstract}

Raw data sizes are growing and proliferating in scientific research, driven by the success of data-hungry computational methods, such as machine learning. The preponderance of proprietary and shoehorned data formats make computations slower and make it harder to reproduce research and to port methods to new platforms. Here we present the RawArray format: a simple, fast, and extensible format for archival storage of multidimensional numeric arrays on disk.

The RawArray file format is a simple concatenation of a header array and a data array. The header comprises seven or more 64-bit unsigned integers. The array data can be anything. Arbitrary user metadata can be appended to an RawArray file if desired, for example to store measurement details, color palettes, or geolocation data.

We present benchmarks showing a factor of 2--3$\times$ speedup over HDF5 for a range of array sizes and a speedup of up to 20$\times$ in reading the common deep learning datasets MNIST and CIFAR10.

\end{abstract}


\section{Introduction}

Our requirements for data storage were that the format should be S.A.F.E.: Simple, Archival, Fast, and Extensible.
\begin{itemize}
\item \emph{Simple}: easily implemented, easily audited for bugs, and mirror the data layout in memory for memory mapping and security;
\item \emph{Archival}: contain all of the information needed to read the data contained in the file with minimal external libraries or code available; subject to introspection through widely available command-line tools;
\item \emph{Fast}: to minimize input and output latency and speed up codes that do heavy input/output and minimize the carbon footprint of the increasing amount of data analysis being done in the world; and
\item \emph{Extensible}: easily improved without a committee and without breaking backward compatibility.
\end{itemize}
None of the existing scientific data formats met these requirements.

Hierarchical Data Format v5 (HDF5\footnote{\url{https://www.hdfgroup.org/HDF5/}}) is burdensome for the casual user to install and to manipulate, raising the barrier to reproducible research. HDF5 also doesn't natively support complex floats, which is mind boggling for a scientific data format.
Julia's JLD2 format is also based on HDF5, unfortunately.

Matlab's MAT format is widespread and relatively easy to manipulate, but the older version 5 doesn't support arrays over 2 GB, and later versions are based on HDF5.

The Nearly Raw Raster Data (NRRD) format\footnote{\url{http://teem.sourceforge.net/nrrd/format.html}} is closest to meeting our requirements, but has some problems. First, it has a flawed type specification. By not separating the numeric bits type (int, float, etc.) from the width of the storage (bits per datum), it is less future proof. For example, the newer \texttt{float16} formats popular in deep learning or very large (512-bit) wide AVX data types are already supported out of the box in RawArray without additional programming. If completely new types come along, we can add additional type codes to RawArray without breaking backward compatibility. Second, NRRD is overly complicated because it implements both raw data and text formats, as well as internal compression, which we prefer to handle externally as part of archives.
Nevertheless, NRRD is a strong competitor and may be preferred for some applications.

Numpy's NPY format is quite fast, but not so simple and is not widely implemented in other languages, making it less portable. But it has, depending on your perspective, an advantage over our format in that it can ``pickle'' derived types and store them automatically. With RawArray, the user is responsible for writing an array of derived types (structs/objects) themselves. Since that is usually as simple as passing a pointer to the array of objects and the total size of the array, we don't consider this to be a significant issue.

Our vision for archiving raw data is metadata as human-readable markup (e.g. YAML, JSON, TOML, INI), raw data in RawArray files,
organized by a file system directory structure,
and packaged in the user's preferred archive file format, such as tar, zip, etc.
This combination is simple, proven, and widely available with battle-tested tools on every platform. 

Coincidentally, Microsoft appears to agree: the latest Office document formats (\texttt{.docx}, \texttt{.xlsx}, \texttt{.pptx}) are actually zip archives of a directory structure containing embedded resources and XML files that describe the document.

\section{Format}

\begin{figure}
\hfil {\bf Table 1}: RawArray File Structure. Note that {\tt u64} $\equiv$ 64-bit unsigned integer, {\tt u8} $\equiv$ 8-bit unsigned integer\hfil
\vskip -10pt
$$\vbox{
\halign{
        #\hfil & \tt # \hfil & \hfil # & \hfil #\cr
\noalign{\smallskip\hrule\smallskip}
{\sc Offset (bytes)}\enspace&\omit {\sc Type} & {\sc Length} & {\sc Description}\cr
\noalign{\smallskip\hrule\smallskip}
0   & u64       & 8 & magic number\cr
8   & u64   & 8 & flags\cr
16  & u64   & 8 & element type code\cr
24  & u64   & 8 & element size (bytes)\cr
32  & u64   & 8 & data length (bytes)\cr
40  & u64   & 8 & number of dimensions\cr
48      & u64[] & 8$\times${\tt ndims} & dimension values\cr
\vdots & \vdots & \vdots\ & \vdots \cr
48 $+$ $8\times${\tt ndims} & u8[] & {\tt data\_length} & array data\cr
\vdots & \vdots & \vdots\ & \vdots \cr
48 $+$ 8$\times${\tt ndims} & {\tt u8[]} &      & optional metadata\cr
\quad $+$ {\tt data\_length} & & \cr
\noalign{\smallskip\hrule}
}}$$
\end{figure}

The first part of the header is a magic number that identifies the file as being a RawArray file.
The magic  number for RawArray was chosen as the ASCII representation of the character string ``rawarray'' (7961727261776172$_{16}$).  This also happens to be eight bytes, so that the entire header can cleanly consist of 64-bit unsigned integers.

The second header field is a bit field that specifies options. This is where byte order and optional or future processing differences can be specified, such as compression or encryption.

\begin{figure}
\hfil {\bf Table 2}: Currently supported elemental type codes\hfil
\vskip -10pt
$$
\vbox{\halign{
#\hfil & \hfil#\cr
\noalign{\smallskip\hrule\smallskip}
{\sc Type Code} & {\sc Description}\cr
\noalign{\smallskip\hrule\smallskip}
0 & user-defined struct\cr
1 & signed integer\cr
2 & unsigned integer\cr
3 & floating point (IEEE-754)\cr
4 & complex float (float tuples)\cr
5+ & reserved for future use\cr
\noalign{\smallskip\hrule}
}}$$
\end{figure}

The third and fourth fields specify the elemental data type, specifically the representation code and the number of bytes per datum. 
Elemental sizes are arbitrary, so quad (128-bit) floats are defined by elemental type 3 and elemental size 16, while the new half (16-bit) floats (useful for deep learning) would be type 3, size 2.

The fifth field is the total size in bytes of the data segment.
This is technically redundant, since it could be computed from the elemental size and the array dimensions, but providing it allows a sanity check and simplifies reading and memory mapping at the cost of just 8 bytes of additional file size.

The sixth and seventh fields are the number of dimensions of the array (e.g. 2 for an image) and a vector that describes the shape of the array. The length of the shape vector should match the number of dimensions specified immediately prior.

Finally, the array data is stored in binary format.  Row and column major do not matter, because that is a detail of the implementation language. As long as the data get read and written in the correct order, everything is portable. 

All data is up front in the file in a linear layout immediately after the purely numerical header to facilitate immediate memory mapping with minimal file parsing. This confers a speed advantage on operating systems that support it. Note that this does not preclude the use of stream compression, such as LZ4, because that can be used to pre-compress the data before storage and uncompress the data when reading. (Although user compression will obfuscate the data within the file and prevent command line tools from easily manipulating the file, thus reducing the ``archival'' quality of the data.)

Checksumming was deliberately omitted because it is difficult to checksum a file containing its checksum. Some formats, such as \texttt{tar}, zero out the checksum field and then checksum the rest of the file, but this requires special software that understands the format, so standard command-line checksum tools won't work.  Finally, checksum algorithms come and go, and ideally scientific data formats would be future proof, without having to modify the archive itself, so we would prefer checksums be computed using the tools of the day and stored external to the data file. To compute an external checksum an RawArray file, simple run your local checksum command. For example, on Linux:
\begin{verbatim}
$ md5sum examples/test.ra
1dd9f98a0d57ec3c4d8ad50343bd20cd  examples/test.ra
\end{verbatim}

Time stamping was also omitted because almost all file systems already provide it. Adding a time stamp that changes upon rewrite or access also foils checksums. HDF5 files are difficult to properly checksum for this reason. Two RawArray files are defined as identical if and only if they contain identical contents, no matter when they were created or accessed last.

\section{Example Usage}

\subsection{Applications}

Using RawArray files in C is extremely simple from the user's perspective. If given a RawArray file called \texttt{mydata.ra}, the C code to read the file  into a RawArray struct, modify it, and write it back out would be simply \begin{verbatim} 
ra_t mydata;
ra_read(&mydata, "mydata.ra");
... do some stuff ...
ra_write(&mydata, "mydata.ra");
\end{verbatim}

The process for the Python, Julia, Matlab, and Rust implementations are similar
In Python, Julia, and Matlab, the RawArray object doesn't even need to be wrapped in a derived type because the arrays in the language contain sufficient information to reconstruct it when writing the file. For example, to read an image in Python that has been stored as a RawArray (something will we demonstrate in the benchmarks below), double the first pixel, then write it back out, you'd simply write
\begin{verbatim}
import ra
img = ra.read('airplane.ra')
img[0, 0] *= 2
ra.write(img, 'airplane.ra') 
\end{verbatim}

In some languages, such as C and Rust, run time type polymorphism is strongly discouraged or even impossible. In those languages, the complied code must know what data type you are trying to import before reading the RawArray file. For example, in Rust, writing a float 32 array and then reading it back would look like
\begin{verbatim}
use rawarray::RawArray;
use std::io;
fn main() -> io::Result<()> {
    let vec1: Vec<f32> = vec![1.0, 2.0, 3.0, 4.0];
    let ra: RawArray<f32> = vec1.clone().into();
    ra.write("myarray.ra")?;

    let vec2: Vec<f32> = RawArray::<f32>::read("myarray.ra")?.into();
    assert_eq!(vec1, vec2);
    Ok(())
}
\end{verbatim}
Notice the type parameters on the RawArray struct that tells the compiler how to monomorphize the methods. This is not ideal, but in practice doesn't seem to present a problem and allows for better optimization.
 
\subsection{Command Line File Introspection}

To get a better handle on the format of an RawArray file, let's look inside one. If you are on a Unix system or have Cygwin installed on Windows, you can examine the contents of an RawArray file using command line tools. For this section, we will use the \texttt{test.ra} file provided in the \texttt{data/} subdirectory of the main repository and the ubiquitous octal dump command, \texttt{od}.

First, let's pretend you don't know the dimensionality of the array. Then
\begin{verbatim}
> od -t uL -N 48 test.ra
0000000              8746397786917265778              0
0000020              4                                8
0000040              96                               2
0000060
\end{verbatim}
shows the dimension (2) as the second number on the third line. The command is extracting the first 48 bytes and formatting them as 64-bit unsigned integers. The ridiculous number listed first is the magic number indicating that this is an RawArray file. 
A slightly different command illuminates that:
\begin{verbatim}
> od -a -N 16 test.ra
0000000    r   a   w   a   r   r   a   y nul nul nul nul nul nul nul nul
0000020
\end{verbatim}
One can already see the shell scripting potential of this format.
Armed with the knowledge that the array is 2D, we know that the header must be 48 + 2*8 = 64 bytes long. The command to skip the header and view only the data would be:
\begin{verbatim}
> od -j 64 -f test.ra
0000100     0.000000e+00            -inf    1.000000e+00   -1.000000e+00
0000120     2.000000e+00   -5.000000e-01    3.000000e+00   -3.333333e-01
0000140     4.000000e+00   -2.500000e-01    5.000000e+00   -2.000000e-01
0000160     6.000000e+00   -1.666667e-01    7.000000e+00   -1.428571e-01
0000200     8.000000e+00   -1.250000e-01    9.000000e+00   -1.111111e-01
0000220     1.000000e+01   -1.000000e-01    1.100000e+01   -9.090909e-02
0000240
\end{verbatim}
Here we are using \texttt{-j} to skip the first 64 bytes and \texttt{-f} to format the byte data as single-precision floats. Note \texttt{od} doesn't understand complex numbers, but the complex data is stored as real and imaginary float pairs that are contiguous on disk. This means that each line of the output is showing two complex numbers with columns 1 and 3 the real parts and columns 2 and 4 the imaginary parts. Notice that it correctly renders the negative infinity.

These examples show how easy it is to manipulate RawArray files on the command line, which improves the accessibility and archival potential of the format.

\section{Benchmarks}

To demonstrate the efficiency gains of RawArray, we ran a series of benchmarks typical of common scientific computing workloads. All benchmarks were run on a Linux server with a RAID5 array of Seagate IronWolf Pro 7200 RPM SATA drives with 256 MB cache and 6 Gb/s transfer speeds (Model ST10000VN0004-1Z).

The first benchmark compared RawArray to HDF5 in reading and writing arrays of a range of strides. The results are shown in Figs.~1 and 2. First test was the time to write 100,000 tiny vectors of length 10, the second test was reading and writing 10,000 small images of size 10 $\times$ 10, and the third test was reading and writing a single large matrix of dimension 10 $\times$ 100,000. The total data size of all tests were kept the same to tease out the overhead of individual operations. Across all tests, RawArray was consistently 2--3$\times$ faster than HDF5.
\begin{figure}
\begin{center}
\includegraphics[width=0.7\linewidth]{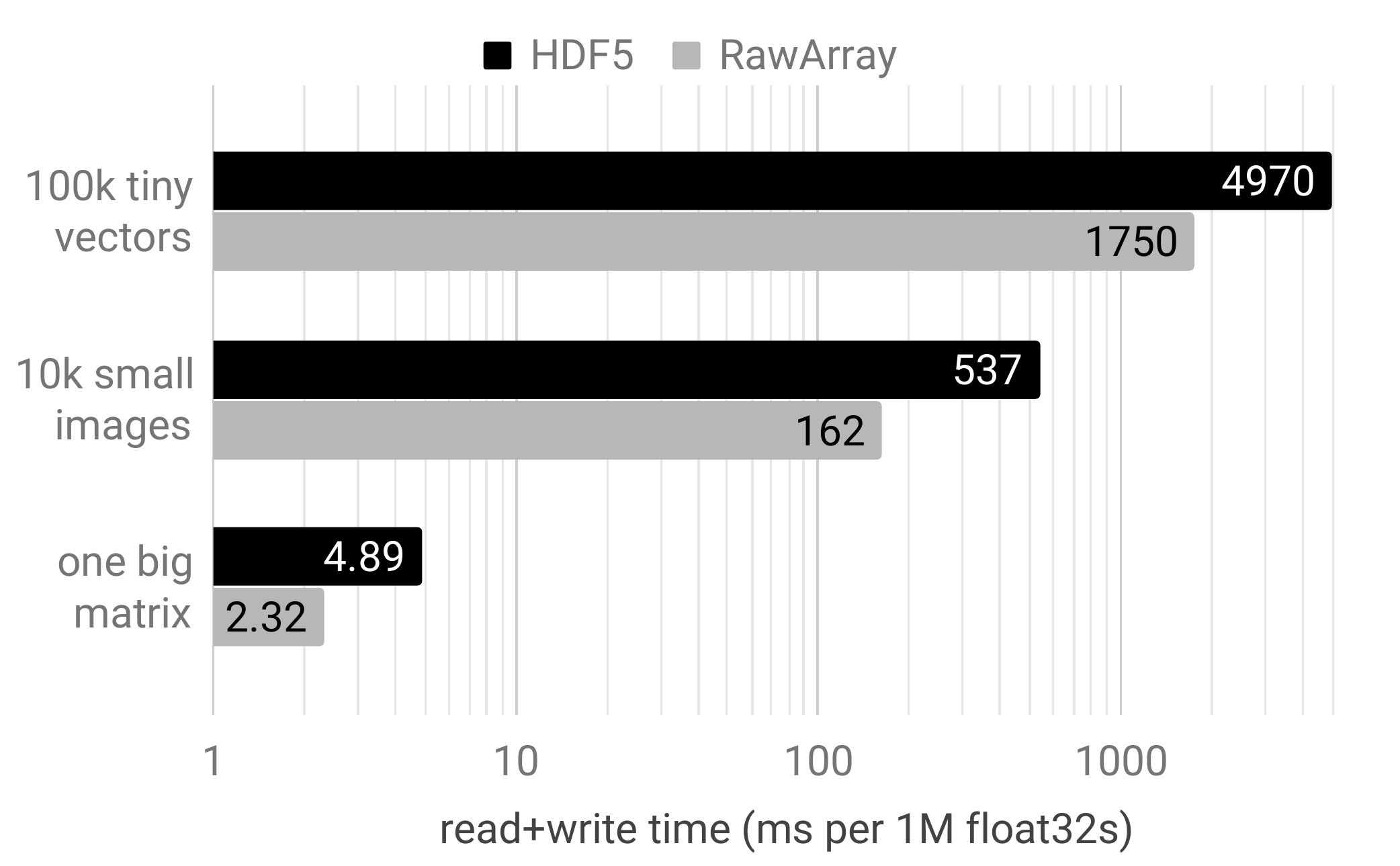}
\caption{Time to write and then read back one million 32-bit floats in a range of configurations. Vectors are length 10, images are $10\times 10$, and the matrix is $10 \times 100,000$. The proposed format was $\sim 2$--$3\times $ faster than HDF5. The total data size read and written was the same; only the number of I/O calls changed. Note the log scale.}
\end{center}
\end{figure}
\begin{figure}
\begin{center}
\includegraphics[width=0.7\linewidth]{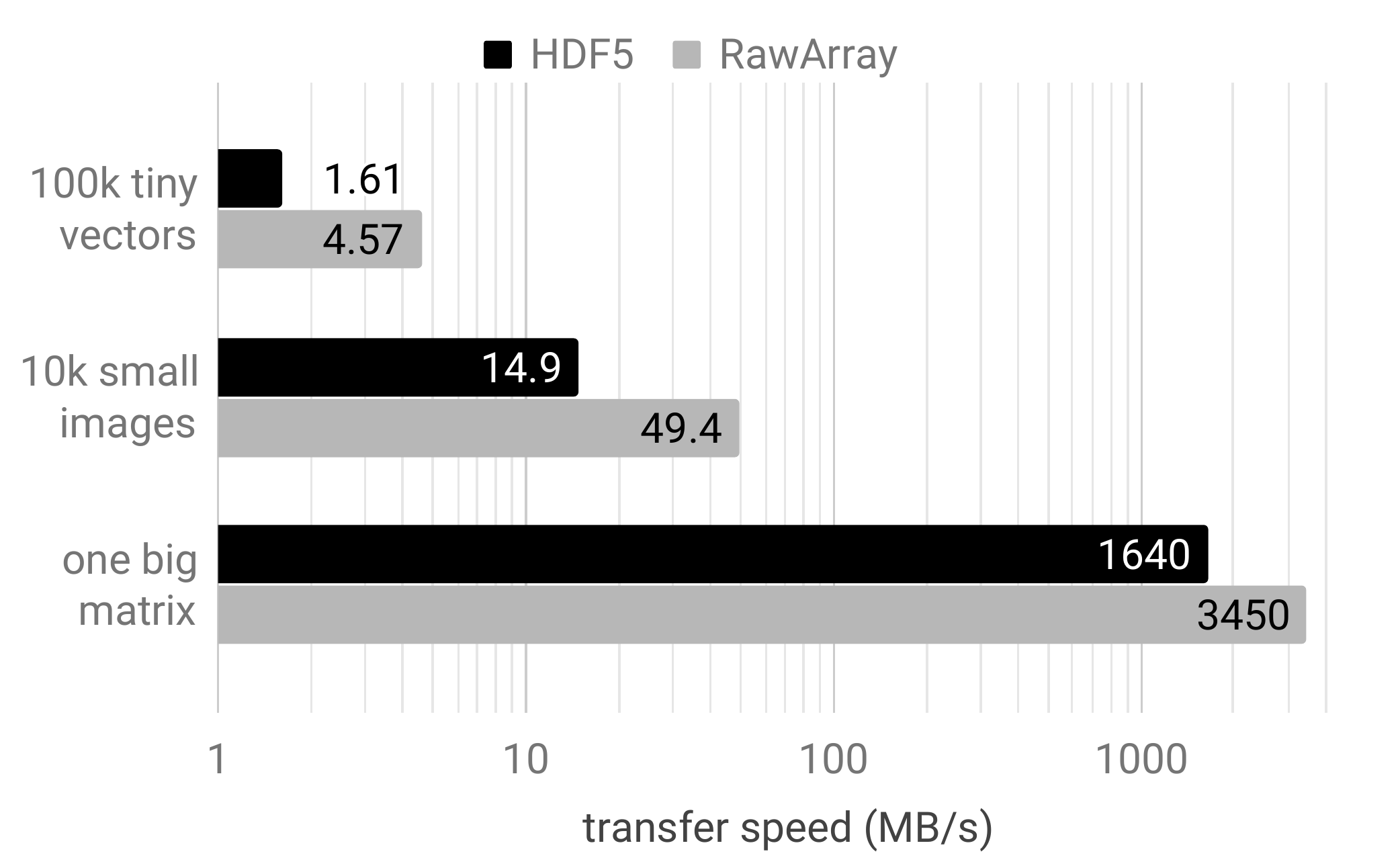}
\caption{Transfer speed to write and then read back one million 32-bit floats in a range of configurations. Vectors are length 10, images are $10\times 10$, and the matrix is $10 \times 100,000$. The proposed format was $\sim 2$--$3\times $ faster than HDF5. Note the log scale.} 
\end{center}
\end{figure}

The second benchmark compared reading images as RawArrays or as the common image format PNG.
In applications with large image datasets, such as deep learning, the images are often stored in common image formats, such as PNG. Furthermore, instead of storing the data in one large archive, it is common for the data to be split among thousands or even millions of small files. 

Here we compare the performance of RawArray and PNG in reading and writing two of the most common deep learning data sets: 
MNIST\footnote{\url{http://yann.lecun.com/exdb/mnist/}} and 
CIFAR10.\footnote{A. Krizhevsky and G. Hinton, Learning multiple layers of features from tiny images, tech. rep., Citeseer, 2009} MNIST comprises 60,000 28$\times$28 pixel grayscale images, while CIFAR10 comprises 60,000 36$\times$36 pixel 3-channel color images. Fig.~3 shows the striking results of using RawArray over PNG. Even though the PNG data is compressed, while the RawArray is uncompressed, so that more data must be read from disk in the RawArray case, our proposed format is 600\% faster reading MNIST and 1800\% faster reading CIFAR. The practical implications are that data loading times during neural network training would be dramatically reduced if the files were read from RawArray files.
\begin{center}
\begin{figure}
\centering
\includegraphics[width=0.7\linewidth]{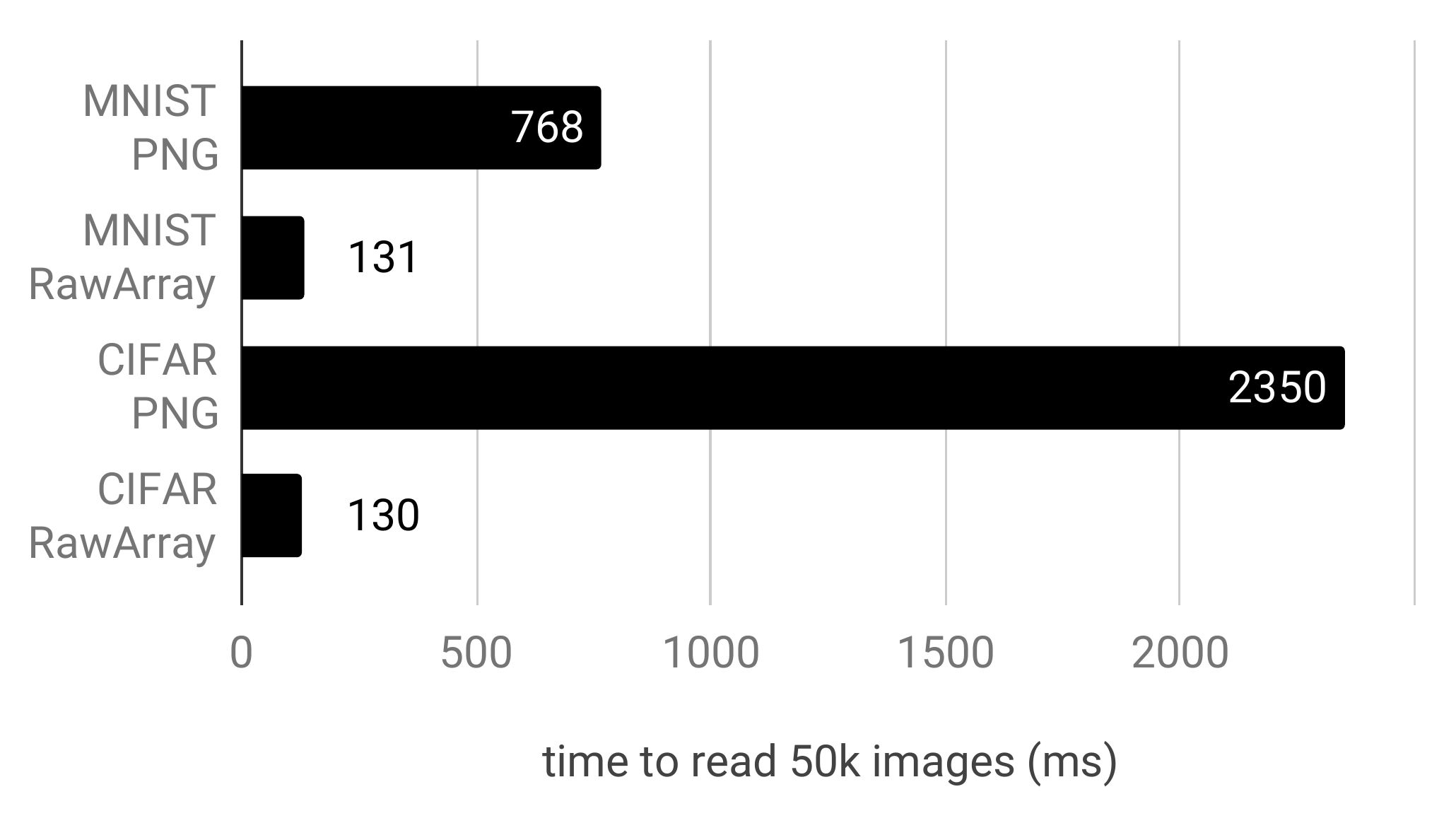}
\caption{Time to read 50,000 small images, typical of the MNIST and CIFAR datasets. The MNIST images are 28$\times$28 single-channel, 8-bit images, while the CIFAR images are $36\times 36$ 8-bit RGB images. Reading the images in RawArray format was 600\% faster for MNIST and 1800\% faster in the CIFAR dataset.}
\label{fig:my_label}
\end{figure}
\end{center}

\section{Discussion and Conclusions}

RawArray is designed to be simple, fast, and highly extensible, while maintaining good qualities for an archival data format.

An example of extensibility is compression or encryption. If at some point in the future, it is decided to add either of these, that can easily be implemented via a new header flag to maintain backward compatibility. Older file versions with the flag naturally set to false will not be affected. Up to 64 different flags could be defined without altering the header structure. 

RawArray could also serve as a file container by using composite types. We think containing files is better left to the file system, however, which is already designed to hierarchically organize heterogeneous objects. But---in theory---you could store RawArray files in a RawArray file, so the format is fractal in that sense. This would be implemented as a user-defined type, where the type is a struct that matches the layout of RawArray file.

RawArray currently handles file sizes up to 18 exabytes ($2^{64}$), but the implementation could be adapted in the future to go beyond that without modification to the format and still retain backward compatibility by exploiting the variable-length list of dimensions. 

Reference implementations exist currently for C, Python, and Matlab\footnote{\url{https://github.com/davidssmith/ra}}, as well as  Julia\footnote{\url{https://github.com/davidssmith/RawArray.jl}} and Rust\footnote{\url{https://github.com/davidssmith/rawarray-rust}}. Contributors and beta testers are welcome, as are helpful suggestions.

\pagebreak

\end{document}